\documentclass[a4paper, 9pt, twocolumn]{article}

\usepackage{geometry}
\usepackage{graphicx}
\usepackage{amssymb}
\usepackage{amsthm}
\usepackage{epstopdf}
\usepackage{mathtools}
\usepackage[shortlabels]{enumitem}
\usepackage{dsfont}
\usepackage{cuted}
\usepackage{sidecap}
\usepackage[usenames,dvipsnames]{color}
\usepackage[thinspace,mediumqspace,Grey,squaren]{SIunits}
\usepackage{bbm}
\usepackage[T1]{fontenc}
\usepackage{authblk}
\usepackage[font=small,labelfont=bf ,textfont={small, sf}]{caption}
\usepackage{titlesec}
\titleformat{\subsection}[runin]{\normalfont\bfseries}{\thesubsection.}{3pt}{}

\newcommand \Expect{\mathbb{E}}
\newcommand{\argmax}{\operatornamewithlimits{argmax}}

\begin{document}

\title{Designing Experiments to Understand the Variability in Biochemical Reaction Networks}
\author{Jakob Ruess, Andreas Milias-Argeitis and John Lygeros \thanks{lygeros@control.ee.ethz.ch}}
\affil{Automatic Control Laboratory, CH-8092 Zurich, ETH Zurich, Switzerland}
\date{}

\maketitle

\abstract{
Exploiting the information provided by the molecular noise of a biological process has proven to be valuable in extracting knowledge about the underlying kinetic parameters and sources of variability from single cell measurements. However, quantifying this additional information a priori, to decide whether a single cell experiment might be beneficial, is currently only possibly in very simple systems where either the chemical master equation is computationally tractable or a Gaussian approximation is appropriate.
Here we show how the information provided by distribution measurements can be approximated from the first four moments of the underlying process. The derived formulas are generally valid for any stochastic kinetic model including models that comprise both intrinsic and extrinsic noise.
This allows us to propose an optimal experimental design framework for heterogeneous cell populations which we employ to compare the utility of dual reporter and perturbation experiments for separating extrinsic and intrinsic noise in a simple model of gene expression. Subsequently, we compare the information content of different experiments which have been performed in an engineered light-switch gene expression system in yeast and show that well chosen gene induction patterns may allow one to identify features of the system which remain hidden in unplanned experiments.}

\section*{Introduction}

Quantitative studies of biological systems with mathematical models strongly depend on an appropriate characterization of the underlying system, that is on good knowledge about the underlying mechanisms and kinetic parameters. While extracting such knowledge from averaged cell population data is common practice, it has only recently been realized that also the molecular noise observed in single cell measurements may be a rich source of information about the parameters of stochastic kinetic models \cite{munsky2009listening,Neuert2013,Zechner2012,Singh2012}. Mathematically, the information provided by single cell measurements can be quantified with the Fisher information. To compute the Fisher information for stochastic kinetic models one has to solve the chemical master equation which governs the time evolution of the probability distribution of the underlying process \cite{Gillespie1992}. This is, however, only possible in the simplest cases and even approximation techniques either remain limited to very small systems \cite{munsky2006finite} or are based on strong assumptions \cite{vanKampen2007,Gillespie2000} which are rarely fulfilled in real applications. Consequently, experiments are usually designed based on the intuition of the experimenter, rather than on information theoretic criteria.

A second difficulty in the analysis and design of single cell experiments is that stochastic kinetic models are usually based on the assumption that the same process governs the evolution in all cells of the population. However, in reality cells are different, due to different cell size, local growth conditions, expression capacity or several other factors \cite{Snijder2011, Shahrezaei2008}. This so-called extrinsic variability \cite{Hilfinger2011,Koeppl,Hilfinger2012} makes the process evolve slightly different in each cell, often leading to additional variability in the response of the cell population. Indeed, in many instances noise resulting from such extrinsic variability has been reported to dominate molecular noise \cite{Elowitz2002,Colman-Lerner2005,volfson2005origins}. In such situations methods that assume a homogeneous cell population and attribute all the observed variability to molecular noise may lead to biased results.
Extrinsic variability does not necessarily refer to ``extrinsic to the cell'' but extrinsic to the model developed for a specific process. Some of this variability may be static for the time scales of interest. For example, the number of mitochondria affecting translation \cite{Zechner2012} changes much more slowly than species in signaling and transcription cascades which form the heart of the model. In this case the number of mitochondria can be taken as random but constant in time for the purposes of the model. Some of the extrinsic variability, however, may be due to species which are not included in the model but affect reaction rates and evolve on time scales comparable to those of the reactions of interest \cite{Hilfinger2011,chabot2007stochastic,rosenfeld2005gene}, for instance global regulators affecting transcription of genes of interest.
In theory one should include such species in the model but this is often not practical since it would lead to models of intractable size. A convenient modeling abstraction may then be to include an assumed stochastic process for some of the rates, to serve as a rudimentary abstraction of the complex mechanisms governing the fluctuations of the reaction rates \cite{Shahrezaei2008}.

In this paper we develop a framework for optimally designing single cell distribution experiments which is applicable in the presence of possibly time-varying extrinsic variability and scalable to systems of realistic size. To this end we first demonstrate on systems where one reaction rate is governed by a stochastic differential equation how equations describing the time evolution of the moments of the probability distribution can be derived for systems which are influenced by time-varying extrinsic variability. We then show how the Fisher information can be approximated from the first four moments without the need of any assumption other than a sufficiently large measured cell population. Finally, we embed the approximated Fisher information into an optimization algorithm which returns the most informative experiment for a specified set of model parameters. This allows us to design optimal experiments for identifying specific features of the system.
We demonstrate this by comparing dual reporter and perturbation experiments in a simple model of gene expression where the mRNA production rate is varying according to a stochastic differential equation. Finally, we study the variability in an engineered light-switch gene expression system in yeast. We use our methodology to evaluate the experiments that were performed in \cite{milias2011silico} and show that they strongly differ in the information provided about the unknown parameters. Further, we show that an experiment found by employing our optimal experimental design procedure would lead to far more information than any of the experiments reported in \cite{milias2011silico}.

\section*{Methods}

\subsection*{Moment equations for reaction systems in fluctuating random environments}

The time evolution of the probability distribution of stochastic kinetic models is governed by the chemical master equation (CME). If extrinsic variability is present in a population, the distribution for each cell can be described by a CME which is conditioned on the realizations of the extrinsic variables in each cell. In \cite{Zechner2012} it was shown how population moments can be computed from this conditional CME under the assumption that the extrinsic variables are constant in time.

More generally, assume now that extrinsic variability can be modeled by a reaction parameter $a_t$ governed by a stochastic differential equation of the form
\begin{equation}
da_t = r(\mu_a - a_t)dt + s\sqrt{a_t} dW_t,
\label{eq:SDE}
\end{equation}
where $W_t$ is a standard Brownian motion. This process fluctuates around its mean $\mu_a$, where the mean reversion speed $r$ gives the autocorrelation time of the process and thereby determines the time scale of the rate fluctuation. The noise coefficient $s$ determines the size of the deviations from the mean, whereas the term $\sqrt{a_t}$ prevents the process from taking negative values and is in accordance with the frequently used Langevin approximation for chemical reaction networks \cite{Gillespie2000}. Note that this formulation includes constant extrinsic variability since for $r=s=0$ $a_t$ is constant and distributed according to its initial distribution.

The system which jointly describes the time evolution of the species and the extrinsic variable is a stochastic hybrid system \cite{hespanha2006modelling}. The time evolution of the moments can be computed as
\begin{equation}
\frac{\mathrm{d}}{\mathrm{d}t} \Expect\left[\psi(a_t,x(t))\right] = \Expect\left[(L\psi)(a_t,x(t))\right],
\label{eq:MomentSys}
\end{equation}
where $x(t) = [x_1(t) \cdots x_m(t)]$ is a vector containing the molecule counts of the $m$ species, $\psi: \mathbb{R} \times \mathbb{R}^m \longrightarrow \mathbb{R}$ is chosen such that the left hand side gives the derivative of the desired moment and $L$ is the extended generator of the stochastic hybrid system, given by
\begin{align*}
(L\psi)(a,x):= & \frac{\partial\psi(a,x)}{\partial a} \cdot f(a) + \frac{1}{2} \frac{\partial^2\psi(a,x)}{\partial a^2} \cdot s^2 a\\
& + \sum_{i=1}^K{\left(\psi(a,x + \nu_i) - \psi(a,x)\right)a_i(a,x)},
\end{align*}
where $f(a) = r(\mu_a - a)$, $\nu_i$ are the stoichiometric transition vectors and $a_i(a,x)$ the propensities of the $K$ reactions. Note that the resulting system of moment equations may be non-closed in the sense that the time evolution of the moments of any order depends on moments of higher order. In such cases the moments cannot be computed exactly and approximation techniques have to be used \cite{ruess2011moment,5605237,Ale2013}.

\subsection*{Approximating the Fisher information}

The amount of information about model structure or parameters which can be gained from measurements may be highly dependent on the experimental setup that is chosen \cite{Bandara2009,Busetto2009,Franceschini20084846,Liepe2013,ZechnerCDC}. Carefully planning an experiment reduces experimental effort and resources and may even allow one to answer questions which cannot be answered from unplanned experiments.

Predicting the information about a vector of unknown model parameters $\theta = \left[\theta_1 \cdots \theta_N\right]^T$ that an experimental setup can supply requires computation of the Fisher information matrix $I(\theta)$, whose elements are given by
\begin{equation*}
\left(I(\theta)\right)_{i,j} = \Expect\left[\left(\frac{\partial}{\partial \theta_i}\log{f(Y;\theta)}\right)\left(\frac{\partial}{\partial \theta_j}\log{f(Y;\theta)}\right)\right],
\end{equation*}
where $Y$ is the random variable which is experimentally measured and $f(Y;\theta)$ is its distribution. In stochastic kinetic models the parameter vector $\theta$ typically contains reaction rates and, if extrinsic variability is present, extrinsic parameters such as moments of the extrinsic distribution \cite{Zechner2012} or parameters describing the fluctuations of the extrinsic variables. Population measurements, such as those provided by flow cytometry, can be viewed as a large number of independent samples $Y_1,\ldots,Y_n$ which are drawn from a marginal distribution $f(Y;\theta)$ of the underlying process. The computation of $f(Y;\theta)$, and therefore also the computation of the Fisher information matrix, requires the solution of the CME. As this is usually not available, methods which can be used to approximate the Fisher information matrix using Monte Carlo simulations have been proposed \cite{Rathinam2010,Gunawan2005}. Such approaches are, however, computationally expensive and hardly feasible in most applications.

An alternative approach resorts to a Gaussian assumption on the underlying Markov process \cite{Komorowski24052011,Komorowski2013}. Under this assumption, the sample mean and variance of the measured population form a jointly sufficient statistic. Hence, computation of the Fisher information reduces to solving the differential equations that describe the dynamics of mean and variance of $f(Y;\theta)$ and computing their partial derivatives with respect to $\theta$. There are, however, many systems where a Gaussian assumption is not reasonable \cite{Zechner2012,Arkin01081998}. In such cases the method in \cite{Komorowski24052011} may lead to erroneous results for several reasons. First, the computed means and variances of $f(Y;\theta)$ may be inaccurate. Second, information that can be gained from higher order moments is neglected. And third, the Gaussian approximation implicitly assumes that sample mean and variance provide independent pieces of information, an assumption which is violated for all non-Gaussian distributions \cite{Lukacs1942}. For instance, for a Poisson distribution the sample mean is already a sufficient statistic on its own and the variance adds no new information (see Supporting Information Section S.1.2).

In situations where a Gaussian assumption is not applicable it may still be possible to approximate the information which is provided by sample mean and variance. If the sample size is sufficiently large, the central limit theorem implies that sample mean and variance are approximately jointly Gaussian.
%(for a more detailed discussion see for instance (\cite{Zechner2012})).
For simplicity, assume that there is only one unknown parameter $\theta$ and that only one species is measured (a more general case is treated in the Supporting Information Section S.1.3). The information given by the mean $I_m(\theta)$ and the joint information given by mean and variance $I_J(\theta)$ can then be approximated using the special form of the Fisher information for multivariate Gaussian random variables (see Supporting Information Sections S.1.1 and S.1.3), which results in

\begin{align}
I_m(\theta) &\approx \tilde{I}_m(\theta) = n\frac{\left(\frac{\partial \mu_1}{\partial \theta}\right)^2}{\mu_2}, \\
I_J(\theta) &\approx \tilde{I}_J(\theta) = \tilde{I}_m(\theta) + n\frac{\left(\mu_2 \frac{\partial \mu_2}{\partial \theta} - \frac{\partial \mu_1}{\partial \theta}\mu_3\right)^2}{\mu_2^2\left(\mu_4 - \mu_2^2\right) - \mu_2 \mu_3^2},
\label{eq:Bounds}
\end{align}
where $n$ is the size of the sample, $\mu_1$ denotes the mean and $\mu_k, k = 2,...,4$ the central moments up to order 4.

These formulas are valid for any distribution which satisfies the requirements of the central limit theorem. Further, it can be shown \cite{Jarret1984} that $\tilde{I}_m(\theta)$ and $\tilde{I}_J(\theta)$ provide lower bounds on the information of the whole sample. For a Gaussian distribution, since $\mu_3 = 0$ and $\mu_4 = 3\mu_2^2$, $\tilde{I}_J(\theta)$ reduces to the correct expression for the complete information. For a Poisson distribution, since $\mu_1 = \mu_2 = \mu_3$, $\tilde{I}_J(\theta)$ reduces to $\tilde{I}_m(\theta)$, which gives the correct expression for the complete information.

\subsection*{Designing optimal experiments}

The goal of experimental design is to find the experiment which is optimal according to some criterion reflecting information about the unknown parameters. The most frequently used criteria are $D$-optimality, $A$-optimality and $E$-optimality which correspond to maximizing the determinant, minimizing the trace of the inverse and maximizing the minimal eigenvalue of the Fisher information matrix, respectively. In biological applications, especially in models which include extrinsic and intrinsic variability, it may be desirable to design experiments which are targeted to specific parameters or to subsets of the parameter set. For instance one might want to estimate the kinetic parameters of the model as well as possible, despite the presence of extrinsic noise. Or conversely, one may be more interested in understanding the extrinsic variability only. An optimality criterion targeted to such questions is $D_s$-optimality \cite{Hunter1969,Walter1990}. It is based on partitioning the parameter vector $\theta = [\theta_1 \ \theta_2]^T$ in parameters of interest $\theta_1$ and nuisance parameters $\theta_2$. The experiment, which allows one to obtain the confidence region with minimum volume for $\theta_1$ can then be found by maximizing the determinant of
\begin{align}
I_s(\theta) &= I_{11}(\theta) - I_{12}(\theta) I_{22}(\theta)^{-1} I_{21}(\theta), \ \mbox{where} \nonumber\\
I(\theta) &= \left[\begin{array}{cc}I_{11}(\theta) & I_{12}(\theta) \\ I_{21}(\theta) & I_{22}(\theta)\end{array}\right],
\label{eq:Information}
\end{align}
$I_{11}(\theta)$ and $I_{22}(\theta)$ are the information matrices for $\theta_1$ and $\theta_2$, respectively, and $I_{12}(\theta)$ and $I_{21}(\theta)$ give the cross terms between $\theta_1$ and $\theta_2$.

The computation of $I_s(\theta)$ requires knowledge of the true parameters $\theta$, which are not available. This difficulty can, for instance, be overcome by evaluating the information at some initial estimate $\hat{\theta}$. If, however, this initial estimate differs significantly from the true parameter vector, the resulting experiment may be far from optimal. This is especially important for biological applications where initial estimates, if available at all, usually involve large uncertainties. Here we chose an approach which includes the uncertainty of the initial estimate in the form of a prior distribution $\pi(\theta)$ and computes the expected information with respect to $\pi(\theta)$ \cite{Pronzato1985,Chaloner1989} (for an overview of other methods see Supporting Information Section S.4).
The corresponding optimal experiment $e^{*}$ can then be obtained by solving the following optimization problem:
\begin{equation}
e^{*} = \argmax_{e \in \mathcal{E}}{\left\{\Expect_{\theta}\left[\det{I_s(\theta,e)}\right]\right\}},
\label{eq:Optimization}
\end{equation}
where the expectation is taken over $\theta \sim \pi(\theta)$, $I_s(\theta,e)$ is the information matrix for experiment $e$ evaluated at $\theta$ and $\mathcal{E}$ is the space of possible experiments.
We can now state a procedure for designing optimal experiments for the estimation of parameters of stochastic kinetic models from measurements of a heterogeneous cell population:\\%[0.1cm]

\definecolor{framecolor}{rgb}{.7,1,.7}
\definecolor{boxcolor}{gray}{.9}
\begin{center}
\fboxrule0.6mm
\fcolorbox{black}{boxcolor}{
\parbox{8.2cm}{
\textbf{The optimal experimental design procedure}\\
\begin{itemize}
\item Include extrinsic variability by taking reaction rates as stochastic processes as in Eq.[\ref{eq:SDE}].
\item Derive the differential equations for the required moments using Eq.[\ref{eq:MomentSys}]. If the moment equations are non-closed and cannot be solved exactly use an approximation method \cite{ruess2011moment,5605237}.
\item Solve the differential equations to compute the moments and their partial derivatives with respect to the parameter vector as functions of $\theta$ and $t$.
\item Choose a vector $\theta_1$ of parameters of interest and specify the distribution $\pi(\theta)$ according to prior uncertainty about $\theta$.
\item Solve the optimization problem Eq.[\ref{eq:Optimization}], where the total information $I_s(\theta,e)$ is replaced by the approximated information of sample mean and variance according to Eq.[\ref{eq:Bounds}].
\end{itemize}
}
}
\end{center}
\vspace{0.3cm}
Some comments on practical applicability of this procedure are given in the SI Appendix Section S.1.4.

This optimal experimental design procedure can be performed in iterations with experiments. Starting from some prior distribution $\pi(\theta)$, the computations lead to optimal experiments that yield data which can be used in a parameter inference scheme \cite{Zechner2012} to compute posterior distributions. These can then in turn serve as new prior distributions for the computation of a new optimal experiment. This can be continued until the uncertainty about the parameters has been sufficiently reduced.

\section*{Results}

\subsection*{In silico study of a gene expression system}

We demonstrate the proposed experimental design framework on a simple example of gene expression. The model we consider consists of the two species mRNA (M) and protein (P) and the four reactions
\begin{align*}
\begin{array}{cccc} \emptyset & \stackrel{a_t \cdot g(t)}{\xrightarrow{\hspace*{1cm}}} & M \\[0.1cm]
 M & \stackrel{b}{\xrightarrow{\hspace*{1cm}}} & \emptyset \\[0.1cm]
 M & \stackrel{c}{\xrightarrow{\hspace*{1cm}}} & M + P \\[0.1cm]
 P & \stackrel{d}{\xrightarrow{\hspace*{1cm}}} & \emptyset, \end{array}
\end{align*}
where $b=0.03, c=0.5, d=0.04$ and $a_t$ is varying according to a stationary stochastic process of the form Eq.[\ref{eq:SDE}]. The dynamics of the moments of order up to two of $M$ and $P$ are then completely determined by the mean reversion speed $r$ and mean $\mu_a$ and variance $V_a$ of the stationary distribution of $a_t$ (see Supporting Information Section S.2.1). Here we assume that the values of these parameters are $\mu_a = 0.5$, $V_a = 0.1$ and $r = 0.005$. We further assume that the gene can be switched between an on state (where $g(t)=1$) and an off state (where $g(t)=0$) using some external input, for example by adding different nutrients in nutrient shift experiments \cite{Menolascina2011}, by adding salt to induce the osmotic stress response \cite{Batt2012}, or with light pulses \cite{milias2011silico}. Further, throughout this section we assume that it is known that no molecules are present at time $t=0$ (loosely speaking the gene has been off for some time at the start of the experiment) and that the degradation rates $b$ and $d$ are known, whereas $\mu_a$, $V_a$, $r$ and $c$ have to be determined from the measurements. Further, for simplicity, all the computations of this section are performed locally at the ``true'' parameter values and prior uncertainty about the parameters is not included.

In the following we compare four experimental methods in terms of the information they can provide about the unknown parameters. For all methods we assume that the experiments are limited to a time length of $t=300$ time units and that at most ten measurements of the protein distribution are taken during that time. The first two methods we consider are standard time course experiments, where the gene is switched on only at time zero and measurements are taken in regular time intervals (every $30$ time units). In the first method (referred to as unplanned experiments) we assume that a single reporter protein is measured, whereas in the second method (unplanned dual reporter experiments) an identical copy of the gene is added to the cells, such that a second reporter protein which is conditionally independent of $P$, given the history of $a_t$, can be measured \cite{Elowitz2002,Swain2002,Bowsher2012}. These two methods are compared to more sophisticated experiments where informative gene switching patterns and measurement times were identified using our experimental design framework. Thereby, the searches for the most informative experiments were performed using a Markov chain Monte Carlo-like algorithm, where we first searched for the optimal gene switching pattern for equally spaced measurement times and then sequentially placed the measurement times where they yielded maximal information for the previously found gene switching pattern. More details on this algorithm are given in the Supporting Information Section S.2.3. Again, we consider both single and dual reporter experiments (referred to as optimal perturbation and optimal dual reporter experiments, respectively). Figure 1 gives the resulting optimal perturbation experiments targeted at the parameters $r$ and $V_a$ (results for the remaining parameters and results for the optimal dual reporter experiments are given in Figures S.2 and S.3 in the Supporting Information). It can be seen that different perturbations and measurement times are optimal for identifying different parameters.

\begin{table*}[!ht]
	\centering 
	\caption{\textbf{Comparison of different experimental approaches.} Information normalized by the sample size $n$. Rows: Information for different parameters of interest. Columns: Information which can be gained by different experimental approaches. Computations corresponding to unplanned experiments were performed with the gene being switched on only once at time zero and equally spaced measurement times. Optimal experiments include optimal gene switching patterns and sequentially placed measurement times (see Supporting Information Section S.2.3).}
	\begin{tabular*}{\hsize}{@{\extracolsep{\fill}}lcccc} 
							            &Unplanned experiment			&Unplanned dual reporter	&Optimal perturbations 	&Optimal dual reporter		\cr \hline
		$\mu_a$ 					    &0.0037										&10.3106 									&4.4499		        		  &10.8805								  \cr 
		$V_a$ 					      &0.0185										&18.6882									&12.8646								&36.9019									\cr 
		$c$ 					        &0.0037										&11.3211									&4.7267									&12.4333									\cr 
		$r$									  &48.5647									&271.5223									&515.6095					  	  &975.4253									\cr \hline 	 
	 \end{tabular*}
\end{table*}

\begin{figure}[!ht]
\centering
\includegraphics{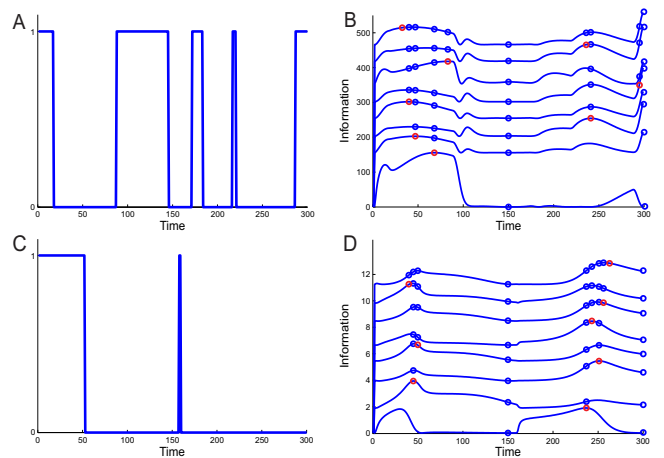}
\caption{\textbf{Optimal perturbation experiments for the parameters $r$ (A and B) and $V_a$ (C and D).} (A,C) Gene switching patterns found by employing the experimental design procedure. A value of one corresponds to the gene being switched on, a value of zero to the gene being switched off. (B,D) Measurement times which attain the highest information for the gene switching patterns of A and C, respectively. Information is normalized by the sample size $n$. In both panels the first two measurement times were placed at $t = 150$ and $t = 300$ (blue circles in the bottommost curve). The bottommost curve shows the information which would be obtained by placing a third measurement at different time points between t=0 and t=300. The measurement is then placed at the time point where the information is maximal (indicated by the red circle on the bottommost curve) and the procedure is repeated for the next measurement. From bottom to top measurements are being placed one at a time (red circle on the corresponding curve) and the total information increases. The blue circles show the location of the measurements that were already placed.}
\label{BestInputs}
\end{figure}

The results of the comparison of the four methods are summarized in Table 1. From the first column we see that the information which can be gained from unplanned experiments is very small. This indicates that the parameters may be practically unidentifiable.
Unplanned dual reporter experiments, on the other hand, lead to much more information (second column in Table 1) and appear to be suitable for identifying both the extrinsic and the intrinsic parameters of the system. The information of the optimal perturbation experiments is given in the third column of Table 1. It can be seen that, compared to dual reporter experiments, more information is obtained for the parameter $r$, whereas dual reporter experiments lead to more information for $\mu_a$, $V_a$ and $c$. Hence, depending on the objective of the study, different experimental strategies are preferable. The fourth experimental method, which combines optimal perturbations and dual reporters, naturally leads to the most information for all parameters (fourth column in Table 1). Note, however, that the increase in information compared to unplanned dual reporter experiments is very small if $\mu_a$ or $c$ are the parameters of interest, which indicates that additionally perturbing the system may not be worth the effort.

Finally, we also computed the information for the first two methods under a Gaussian assumption. Our results (Supporting Information Table S.1) show that for many objectives a Gaussian assumption leads to information estimates which are overly optimistic. This is most likely due to the independence assumption of sample mean and variance which is implicitly imposed by a Gaussian approximation.

\subsection*{Characterizing extrinsic noise in a light-induced gene expression system}

Next we study a light-switch gene expression system which has been engineered in yeast \cite{milias2011silico}. The authors used a light responsive module to initiate transcription by shining red light on the yeast culture and to terminate transcription by shining far-red light. They then proposed a control scheme to regulate the mean amount of protein in the population. The development of more sophisticated control schemes (for example, to allow one to also control the protein variance) requires a proper characterization of the sources of variability in the system. To this end our framework can serve to compare the utility of different experiments and ultimately to design the experiments which are optimal for the characterization of the different noise sources. 
We thus developed a stochastic version of the model in \cite{milias2011silico} with time varying extrinsic variability. This introduced four additional parameters which were not identified in \cite{milias2011silico}. To characterize uncertainty about these parameters we chose independent uniform prior distributions and computed the expectation of the information which is provided by the experiments reported in Figure 1 of \cite{milias2011silico} about each of the additional parameters according to Eq.[\ref{eq:Information}]. Thereby, the remaining parameters which were already identified in \cite{milias2011silico} were fixed to their known values (see Supporting Information Section S.3.1 and Section S.3.2). The results are shown in Tables S.2 and S.3 in the Supporting Information. Which experiment is best again depends on the objective. For instance, the experiment where a red light pulse is applied at the beginning and a far-red light pulse after 30 minutes is best to identify the protein production rate but worst for all other parameters. This is most likely due to the fact that if the gene is switched off, the mRNA production rate is set to zero and the parameters describing this rate do not influence the dynamics anymore.

Further, our results show that even though the experiments in Figure 1c of \cite{milias2011silico} were performed over a longer time and contain more measurements than the experiments in Figure 1e of \cite{milias2011silico}, the experiments in Figure 1e provide much more information about the mean reversion speed $r$ of the extrinsic fluctuations.
This suggests that experiments where the gene is expressed for short time intervals with silent periods in between could allow one to determine $r$ and thus to test whether extrinsic variability can be assumed to be constant or whether a stochastic process description as used in this paper is required.
Our results in Section S.3.3 in the Supporting Information indicate that indeed, contrary to other experiments, the variance measured in the experiments in Figure 1e in \cite{milias2011silico} cannot be explained very well by a model with constant extrinsic variability.

Finally, we also employed our experimental design framework to search for experiments which carry high information about $r$. The light-pulse pattern and measurement times we determined with our method are shown in Figure 2. They lead to an experiment which carries close to four times more information than any of the experiments in \cite{milias2011silico} suggesting that our experimental design framework can be a valuable tool for characterizing variability in this system.

\begin{figure}[!ht]
\centering
\includegraphics{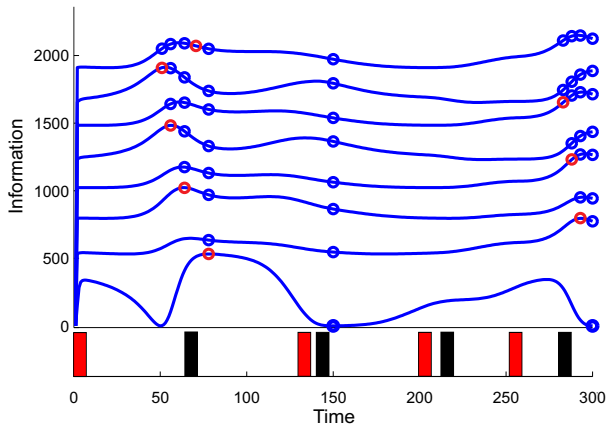}
\caption{\textbf{Results of experimental design targeted at the parameter $r$ for the light-switch gene expression system.} The optimal light-pulse pattern is shown at the bottom of the figure. Red rectangles correspond to red light pulses, black rectangles correspond to far-red light pulses. The upper part of the figure shows the optimal measurement times, where the interpretation is the same as in Figure 1.}
\label{OptimResultsLightSwitch}
\end{figure}

\section*{Discussion}

While knowledge about biological mechanisms is constantly growing, our understanding of the stochasticity of biological systems and its influence on system dynamics remains rather limited. Many different sources of variability may play a role and neglecting any one of them may lead to substantial errors. For instance, neglecting extrinsic variability in the gene expression model and using the ideas in \cite{PhysRevLett.97.168302} to quantify protein burst size and frequency from measurements of the protein distribution would lead to conclusions which are more than an order of magnitude wrong (see Supporting Information Section S.2.2). Allowing reaction rates to vary between individuals in a cell population, either by taking them as unknown constant random variables or by allowing them to be governed by stochastic differential equations, offers a way to incorporate extrinsic variability in a model and enables model-based studies of heterogeneous cell populations.
We showed how the information about unknown parameters of such models which is provided by means and variances of measured populations can be approximated by solving a set of ordinary differential equations. The derived formulas apply to any kind of system with the only assumption that the measured population is of sufficient size for the application of the central limit theorem.
This opens up the possibility to pose many interesting questions: do the measurements contain enough information to separate the different noise sources? How much information can be gained by measuring the variance in addition to the mean? And most importantly: what is the most informative experiment?
By means of examples we demonstrated that unplanned experiments may not contain enough information to separate the different sources of variability and that designing experiments based on intuition alone may not be sufficient. For instance, placing all the measurements either very early or very late in the experiment turns out to be optimal for identifying the mean reversion speed in our examples (see Figures 1 and 2), but appears very unintuitive at a first glance.

Our results (Table 1 and Supporting Information Figure S.2) show that the optimal experiments are highly dependent on the chosen objective. In some cases introducing a dual reporter yields high information, in other cases perturbing the system with input stimuli is preferable. A study of the system using the experimental design framework presented in this paper allows a comparison of the different experimental approaches and enables one to choose the approach which is most likely to be successful for the given objective. The resulting experiments can in turn be used to refine the model and to update the parameter estimates, giving rise to an iterative procedure of successive rounds of computations and experiments.

In the light-switch gene expression system the computation of the information contents of the different experiments shows that perturbing the system with different light pulse sequences can highlight different features of the system. This suggests that well chosen gene induction patterns may allow one to uncover features of the system which remain hidden in unplanned experiments. For instance
Figure S.4 in the Supporting Information suggests that temporal fluctuations in the extrinsic variable may play a role for this system.
Perturbing a system with light pulses to understand the variability may seem to be limited to this specific engineered system. However, a similar strategy could also be employed by exploiting naturally occurring biological mechanisms. For example, in \cite{Zechner2012} the authors studied gene expression in yeast in response to osmotic pressure. Different salt concentrations lead to different residence times of the signaling molecule Hog1 in the nucleus and thereby created different input signals to the downstream gene expression system. In that system multiple subsequent salt stresses, which can for instance be implemented using a micro-fluidic device as in \cite{Batt2012}, could serve as the equivalent of the multiple light-pulses used in this paper and might give further insights into the specific nature of the system.

\subsection*{Acknowledgements}
The work was supported in part by the European Commission under the project MoVeS.

\subsection*{Author Contributions}

J.R. and J.L. designed research. J.R. developed the method and performed the computations. A.M. assisted with supplementary calculations. J.R. and J.L. wrote the paper.

\subsection*{Supporting Information}
A PDF document with supporting discussions and results is available at http://control.ee.ethz.ch/$\sim$jakorues/DesignSI/.

\end{document}